\documentclass[journal = ancac3,manuscript = article]{achemso}

\usepackage{amsmath}
\usepackage{graphicx}

\usepackage[version=3]{mhchem}

\author{Rodrigo G. Amorim}
\email{rodrigo.amorim@physics.uu.se}
\affiliation
{Division of Materials Theory, Department of Physics and Astronomy, Uppsala University,\\Box 516, SE-751 20 Uppsala, Sweden}
\author{Ralph H. Scheicher}
\email{ralph.scheicher@physics.uu.se}
\affiliation
{Division of Materials Theory, Department of Physics and Astronomy, Uppsala University,\\Box 516, SE-751 20 Uppsala, Sweden}

\title[Silicene as a new ultrafast DNA sequencing device]{Silicene as a new ultrafast DNA sequencing device}

\abbreviations{IR,NMR,UV}

\begin{document}
\maketitle

\begin{abstract}
Silicene, a hexagonal buckled 2-D allotrope of silicon, shows potential as a platform for numerous new applications, including bio-sensing. We have used density functional theory, incorporating corrections to account for van der Waals interaction, to investigate the adsorption of individual nucleobases on silicene. Our study sheds light on the stability of this novel nano-bio system, its electronic properties and the pronounced effects on the transverse electronic transport, i.e., changes in the transmission and the conductance caused by adsorption of each nucleobase, explored by us through the non-equilibrium Green's function method. Intriguingly, despite the relatively weak interaction between nucleobases and silicene, significant changes in the transmittance at zero bias are predicted by us, in particular for the two nucleobases cytosine and guanine. Our findings suggest that silicene could be utilized as an integrated-circuit biosensor as part of a lab-on-a-chip device for DNA analysis.
\end{abstract}

It has long been known that, under standard conditions, graphite is the most stable form of carbon while silicon takes on the diamond cubic crystal structure. This natural circumstance is a direct reflection of the greater stability of the $sp^2$ hybridization in C and of the $sp^3$ hybridization in Si. Nonetheless, it was theoretically predicted \cite{Takeda1994, Fagan2000a, Guzman-Verri2007a, Cahangirov2009} and subsequently experimentally confirmed \cite{Aufray2010, Lalmi2010a, Lin2012a, Jamgotchian2012, Feng2012, Vogt2012, Fleurence2012}, that a 2-D honeycomb lattice like graphene can also be formed by Si, named silicene. Despite some structural similarities of silicene with graphene, the two differ in a couple of crucial aspects: firstly, the most stable structure of silicene is not planar, but buckled. This feature was described already in its original theoretical prediction \cite{Takeda1994, Cahangirov2009} and also observed in experiments \cite{Lin2012a, Vogt2012, Fleurence2012}.
Secondly, the method to obtain silicene is rather different from that used originally for graphene \cite{Novoselov2004}, in that for silicene one cannot apply the exfoliation method since this allotrope of silicon cannot freely exist in nature. Instead, the prevalent method to synthesize silicene is to deposit Si atoms on a substrate in the form of a metal surface.

In the present work, we investigate whether silicene could be used for third-generation DNA sequencing \cite{Branton2008, Fyta2011, Venkatesan2011, Scheicher2012, DiVentra2013, Yang2013, Haque2013}. Graphene has been extensively studied with this application in mind \cite{Postma2010, Merchant2010, Schneider2010, Garaj2010, Nelson2010, Min2011, Cho2011, Bergvall2011, Sathe2011, He2011, Prasongkit2011, Liu2012a, Saha2012, Venkatesan2012, Cheng2012, Qiu2012, Wells2012, Liu2012, Lv2013, SlavenGaraj2013, Li2013, Hui2013, Avdoshenko2013, Prasongkit2013, Jeong2013, Freedman2013, Dekker2013, Traversi2013}. Silicene represents an interesting alternative and might even offer an advantage here over graphene in terms of integrability with existing silicon-based microchip technology. To evaluate its capability for distinguishing the four different nucleobase types of DNA electronically, we investigated the stability, electronic structures and transport properties of silicene with each nucleobase adsorbed on top of silicene. This approach resembles most closely that investigated by the group of K. S. Kim for graphene nanoribbons \cite{Min2011, Cho2011}. Our results reveal that, although the electronic structure is not drastically changed due to the relatively weak interaction, it might be possible to detect the nucleobase type by analyzing subtle changes in the transport properties of the systems.

\begin{figure*}
\hspace{-1cm}
\centering
\includegraphics[scale=0.18]{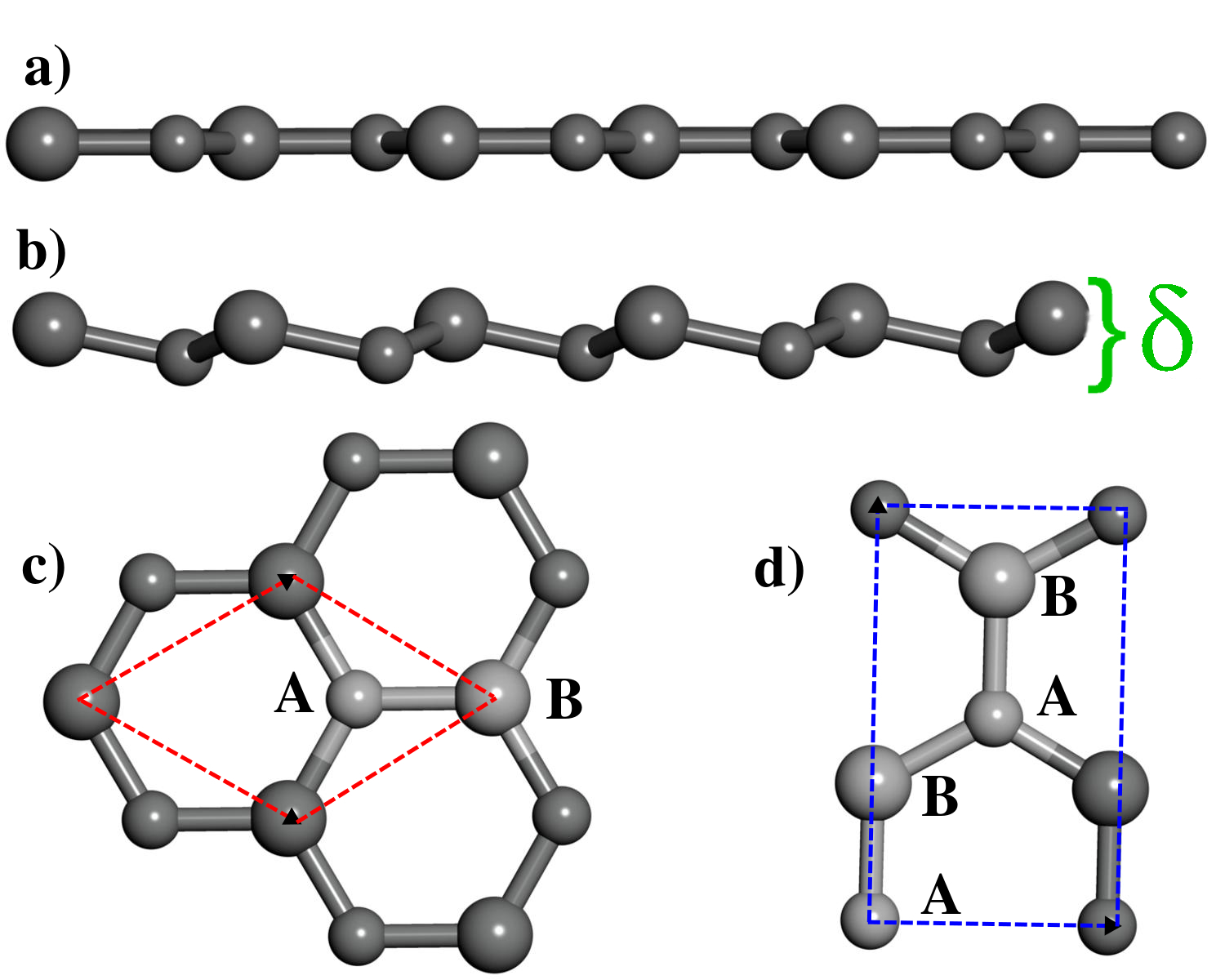}
\caption{Schematics of relaxed supercells for (a) flat and (b) buckled structures. Two choices of unit cells are shown in (c) hexagonal with two Si atoms and (d) parallelepiped with four Si atoms per unit cell. Dashed lines in (c) and (d) represent the boundaries of the respective unit cells and light gray atoms mark the atoms included in each unit cell.}
\label{figure1}
\end{figure*}

\section{Pristine Silicene}


In order to establish the reliability of our computational methods to accurately describe pristine silicene itself, we first carried out a number of benchmark tests concerning the structural and electronic properties of silicene. Figure \ref{figure1} shows the fully relaxed supercell for flat and buckled structures in which $\delta$ represents the vertical distortion of the atomic positions from a perfectly planar geometry. The flat structure is found to be metastable in the sense that a small out-of-plane displacement of the atoms would lead to the sheet becoming buckled upon relaxation. Both the hexagonal and parallelepiped unit cells with two and four atoms, respectively, were considered by us (Figure \ref{figure1}). The relaxed lattice constant, Si-Si distance, and resulting vertical distortion $\delta$ are listed in Table \ref{silicene_structure}, calculated with the exchange-correlation potential GGA (PBE) and with GGA including van der Waals (vdW) corrections. Our results for the Si-Si distance ($2.27-2.31$ \AA) are in excellent agreement with experimental results \cite{Vogt2012} ($2.2 \pm 0.1$ \AA) and with recent theoretical predictions\cite{Cahangirov2009} ($2.25$ \AA). For the proper description of pristine silicene, it is of rather little difference whether vdW corrections are added to GGA or not. However, since a major goal of our present work is to study the physisorption of DNA nucleobases on silicene, a process in which vdW interactions may be crucial, the conclusion from our benchmark results in Table \ref{silicene_structure} is that GGA+vdW does indeed yield accurate results for pristine silicene.

\begin{table}[!h]
\caption{Lattice constant ($a$), bond length ($d_{Si-Si}$), and vertical distortion ($\delta$) of planar and buckled silicene in units of \AA. These fully relaxed structural parameters of pristine silicene were obtained using the exchange-correlation approximation GGA alone and GGA with van der Waals (vdW) corrections added.}


\begin{center}
\begin{tabular}{||l||c|c|c|c|c|c||}
\hline
 &  \multicolumn{3}{c|}{GGA}         & \multicolumn{3}{c||}{GGA+vdW}
\\
\hline
structure &  $a$     &$d_{Si-Si}$& $\delta$&    $a$     &$d_{Si-Si}$&$ \delta$ \\ \hline
 planar     &$3.945$ &$2.27$     &$0.00$   &$3.995$   &$2.30$     &$0.00$    \\
 buckled  &$3.949$ &$2.27$     & $0.46$  &$4.002$   &$2.31$     & $0.51$   \\ \hline
\end{tabular}
\end{center}
\label{silicene_structure}
\end{table}

\begin{figure*}
\hspace{-1cm}
\centering
\includegraphics[scale=0.115]{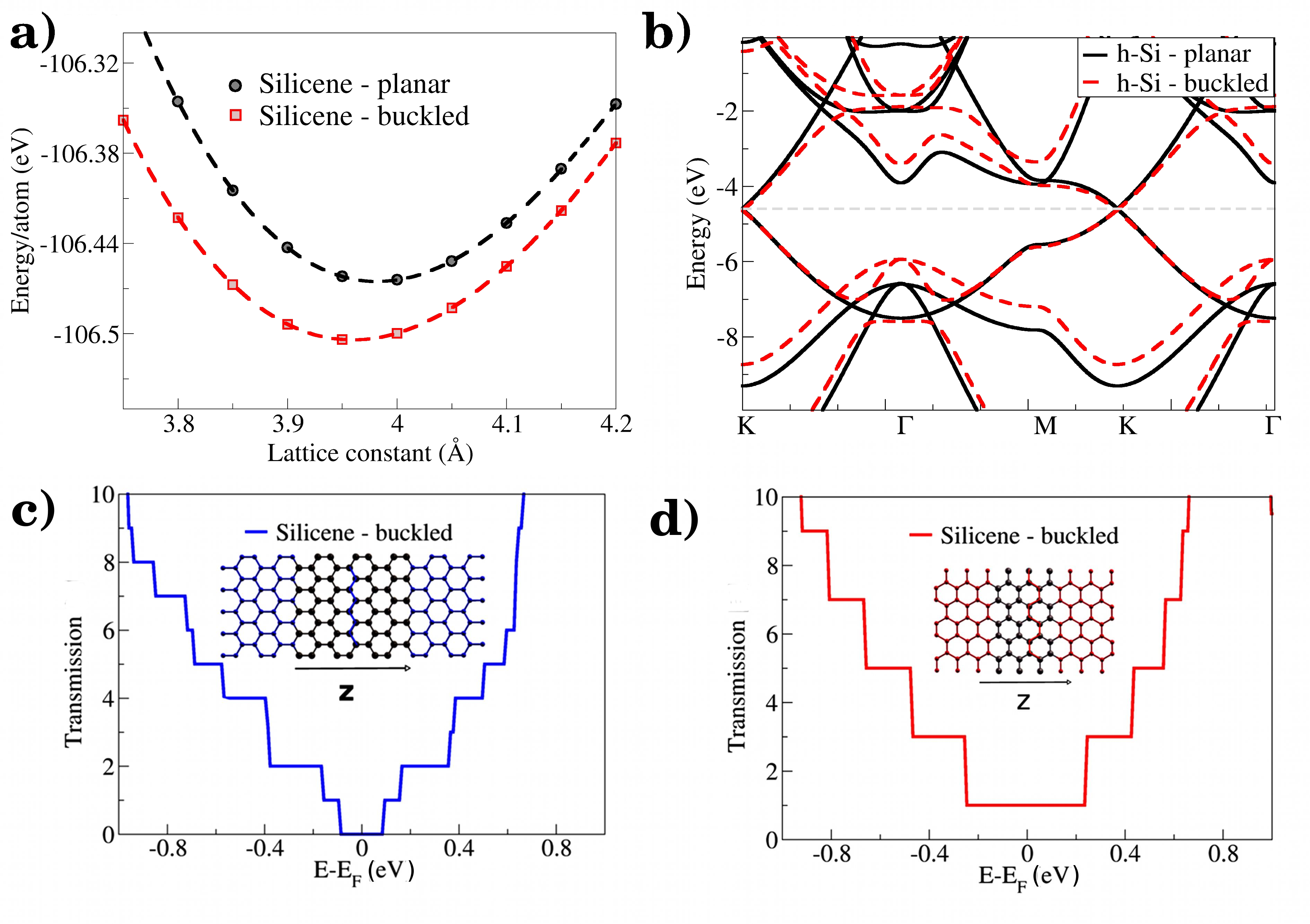}
\caption{Pristine silicene properties are shown: (a) total energy as a function of the lattice constant, illustrating the relatively higher stability of buckled silicene compared to the planar allotrope; (b) band dispersion of planar and buckled configurations; and electronic transport in the $z$-direction perpendicular to (c) zigzag rows and (d) armchair rows of Si atoms.}
\label{pristine_silicene_energetics}
\end{figure*}

We furthermore investigated the energetic properties of pristine silicene, as summarized in Figure \ref{pristine_silicene_energetics}a. It can be seen that the buckled structure is approximately 0.05 eV/atom 
more stable than the planar one, in agreement with theoretical predictions\cite{Cahangirov2009} and the experimental\cite{Vogt2012,Fleurence2012} findings that silicene exists in a buckled configuration. Calculated band structures for the planar and buckled structures are compared in Figure \ref{pristine_silicene_energetics}b. One can note the similarities with the band structure of graphene, inasmuch as we observe the same crossing of the $\pi$ and $\pi^{*}$ bands at $K$ and $K^\prime$ points in the Brillouin zone. The main difference between the electronic band structures of the planar and buckled geometries is a varying shift in energy of the respective bands, with the magnitude of the shift smoothly changing as we move through reciprocal space.

The calculated transmittance shown in Figure \ref{pristine_silicene_energetics}c and \ref{pristine_silicene_energetics}d considers transport in directions perpendicular to zigzag (c) and armchair (d) rows. As we can see, transport perpendicular to the zigzag rows exhibits a gap in the transmittance at $E=E_F$, while transport perpendicular to the armchair rows possess finite normalized transmittance for all energies. This difference is solely due to the computational necessity of using a finite grid of $k$-points for Brillouin zone sampling. If an infinitely dense grid $k$-points could be considered, the electronic transmission curves for both directions would be indistinguishable and the discrete steps would take on an essentially smooth linear shape as is common for Dirac materials.

\section{Equilibrium Geometries}

Having established the basic structural and electronic properties of pristine silicene and their accurate description through the computational methods chosen by us in the present study, we now turn our attention to the main focus of this paper, namely the physisorption of DNA nucleobases on silicene and the consequential effects on the electronic conductance properties of this nano-bio hybrid system.

\begin{figure*}
\hspace{-1cm}
\centering
\includegraphics[scale=0.52]{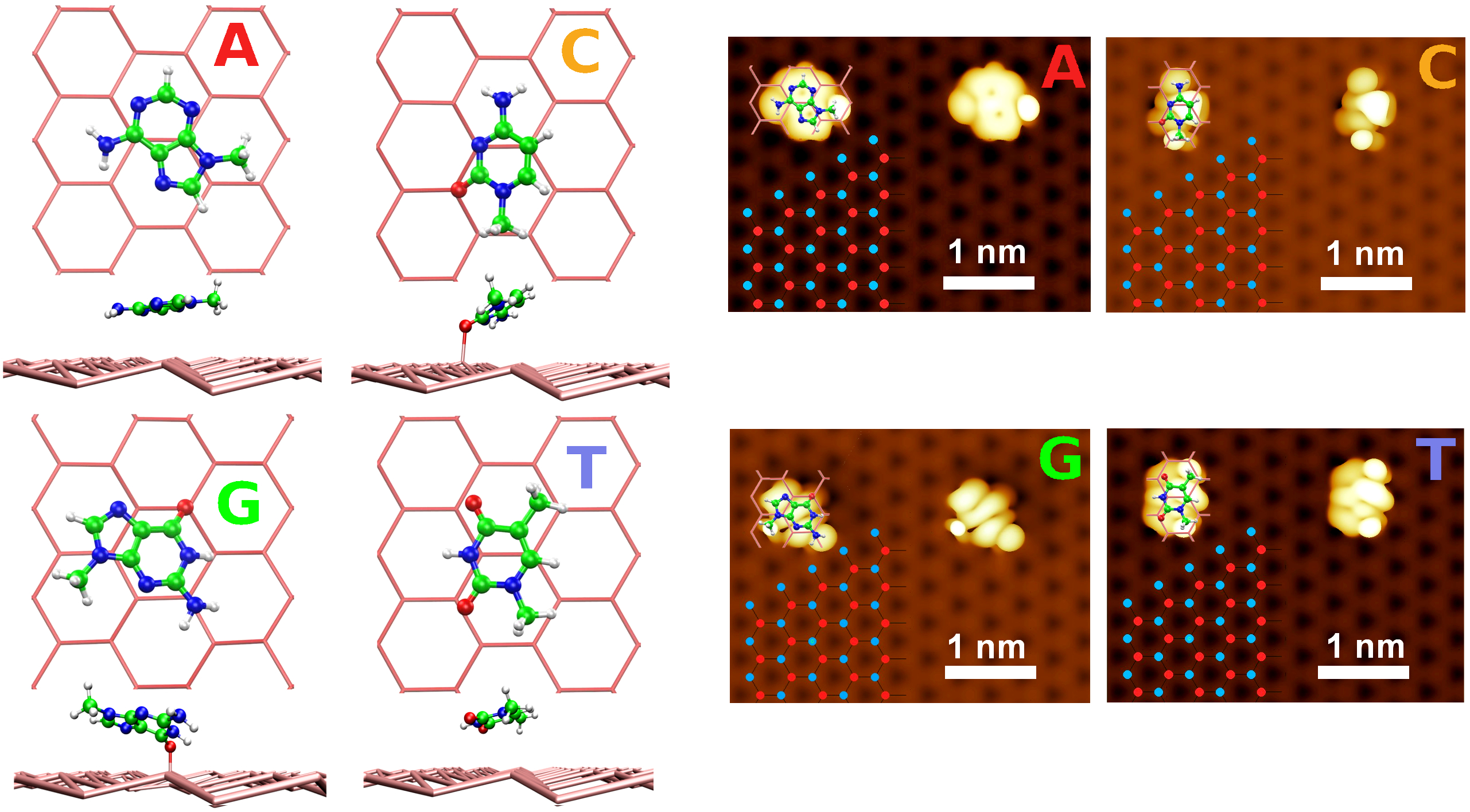}
\caption{The left panel shows top views and side views of the fully relaxed geometries for the four different nucleobases (A, C, G, and T) physisorbed on a buckled silicene sheet. The hexagonal network of Si atoms is drawn in salmon color, C atoms are represented as green spheres, N atoms in blue, O atoms in red, and H atoms in white. The right panel shows the nucleobase fingerprints on silicene as they would appear in an image recorded by a scanning tunneling microscope (STM). The STM images were generated with the $WSxM$ software \cite{Horcas2007} from our DFT calculations using the Tersoff-Hamann approximation for filled states. STM fingerprints for each nucleobase are shown twice for clarity, once with the atomic structure superimposed and once without it. The hexagonal patterns of dots to the lower left in each image is meant to help locating the Si atoms from the buckled silicene sheet which are either extended upwards (blue dots) or downwards (red dots) from the plane.}
\label{bases_on_silicene}
\end{figure*}

The fully relaxed equilibrium geometries of the four DNA nucleobases adenine, cytosine, guanine, and thymine (abbreviated as A, C, G, and T in the following) physisorbed on silicene are shown in the left panel of Figure \ref{bases_on_silicene}. We first note the larger size of the underlying hexagons formed by the silicon atoms compared to the smaller six- and five-membered rings formed by carbon and nitrogen atoms in the nucleobases. This size mismatch is an important difference to similar physisorption processes of DNA nucleobases on graphene and boron nitride studied in the past \cite{Mukhopadhyay2010,Lee2013,Gowtham2007a}.

One feature discernible from the side view of Figure \ref{bases_on_silicene} is that cytosine and guanine have their single oxygen atom located directly above a silicon atom. For thymine with its two oxygen atoms, no such alignment is observed (and for geometrical reasons, it is actually impossible to align both oxygen atoms simultaneously on top of the underlying silicon atoms without inducing drastic deformations in the bond angles or bond lengths of thymine). In terms of the equilibrium distance between nucleobases and silicene, we can distinguish two categories: one category (A and T) in which the closest atomic distance between the two entities is a (non-covalent) Si--H connection with a distance of about 3 \AA\ and a second category (C and G) in which the closest distance is given by a Si--O connection amounting merely to around 2 \AA. These results are quantitatively summarized in Table \ref{tab:distances}. The correlation between equilibrium distance and binding energy is discussed further below.
\begin{table}[!h]
\caption{For each nucleobase (A, C, G, and T) the closest pair of atoms between the silicene sheet (Si) and the base (either H or O) are given along with their respective distance in \AA. The corresponding binding energies were calculated by performing a vertical energy scan (see Figure \ref{vertscan}) and are given in eV as the energy relative to infinitely far separated nucleobase and silicene sheet. Negative binding energies thus indicate stable configurations.}
\begin{center}
\begin{tabular}{|c|c|c|c|}
\hline
nucleobase   & closest atoms &   distance Si-base&  binding energy \\ \hline
A               &   Si-H       &   $3.065$         &   $-0.604$        \\
C               &   Si-O       &   $2.039$         &   $-0.881$        \\
G               &   Si-O       &   $1.990$         &   $-0.774$        \\
T               &   Si-H       &   $3.070$         &   $-0.601$        \\
\hline
\end{tabular}
\end{center}
\label{tab:distances}
\end{table}

\section{STM Images}


In the right panel of Figure \ref{bases_on_silicene} we show through a series of simulated images how the four different nucleobases would appear in a scanning tunneling microscope (STM). To calculate the images, we apply the Tersoff-Hamann \cite{TersoffJandHamann1985} approach, in which the electronic state of the STM tip is modeled by an $s$-orbital and the assumption is made that the sample wave-functions near the tip exhibit only small variations. The resulting tunneling current (or rather differential conductance $dI/dV$) is proportional to local density of states (LDOS) or partial density of charge $\rho(z_0, E_F+V)$ integrated from a specific energy to the Fermi energy ($E_F$), where $z_0$ is the tip height and $E_F$ to $E_F+V$ is the range of energy considered in the LDOS calculation. For filled states, we integrated from $-3$eV to $E_F$ (see partial densities of states, PDOS, in the Supporting Information).

Considering the tip to be at the same height for all STM images (we used the same isovalue for LDOS), it is possible to verify that the silicene sheet in the panels for A and T appears in a darker color compared with those for C and G. The explanation is that the bases A and T are further away from silicene than C and G (see Table \ref{tab:distances}). The brightness of the nucleobases are all similar because the distance (tip-nucleobase) is almost the same for all.

Now we discuss the correlation of the STM images with the equilibrium geometries. For Adenine the final structure (see top and side views in the left panel of Figure \ref{bases_on_silicene}) is less tilted compared to the other nucleobases (C, G, and T), and therefore both rings of this nucleobase are clearly discernible in the STM image. The methyl group (CH$_3$) has one hydrogen pointing up, which is thus the closest atom to the STM tip and thus appears as the brightest spot in this region. The amine group (NH$_2$) does not possess such a bright signature, because it is somewhat lower in height compared to the H atom from the CH$_3$ group. We also identify the two hydrogen atoms bonded to carbon atoms, as two small bulges. For Guanine the equilibrium geometry is tilted (see side view, left panel of Figure \ref{bases_on_silicene}), and because of that, it is more difficult to recognize the rings as in the case of Adenine. However, the methyl group is pointing up and thus has one of its H atoms appear as the brightest spot, similar to the Adenine case. The oxygen atom is pointing down towards the substrate and the corresponding intensity is therefore less bright. Cytosine possesses the same tilted equilibrium geometry as Guanine (see side view, left panel of Figure \ref{bases_on_silicene}) and a similar interpretation is therefore possible here. The main differences are the two hydrogen atoms pointing up at the same side of the molecule. These two hydrogen atoms appear as a large bright spot on the right side of the right side of Cytosine. Finally, for Thymine we observe two bright spots due two CH$_3$ groups. The two oxygen atoms are pointing slightly down and the brightness is therefore less strong for them. These nucleobase fingerprints could be an important guide for experimentalists trying to identify DNA nucleobase adsorbed on silicene.

\section{Binding Energies}

\begin{figure*}
\hspace{-1cm}
\centering
\includegraphics[scale=0.8]{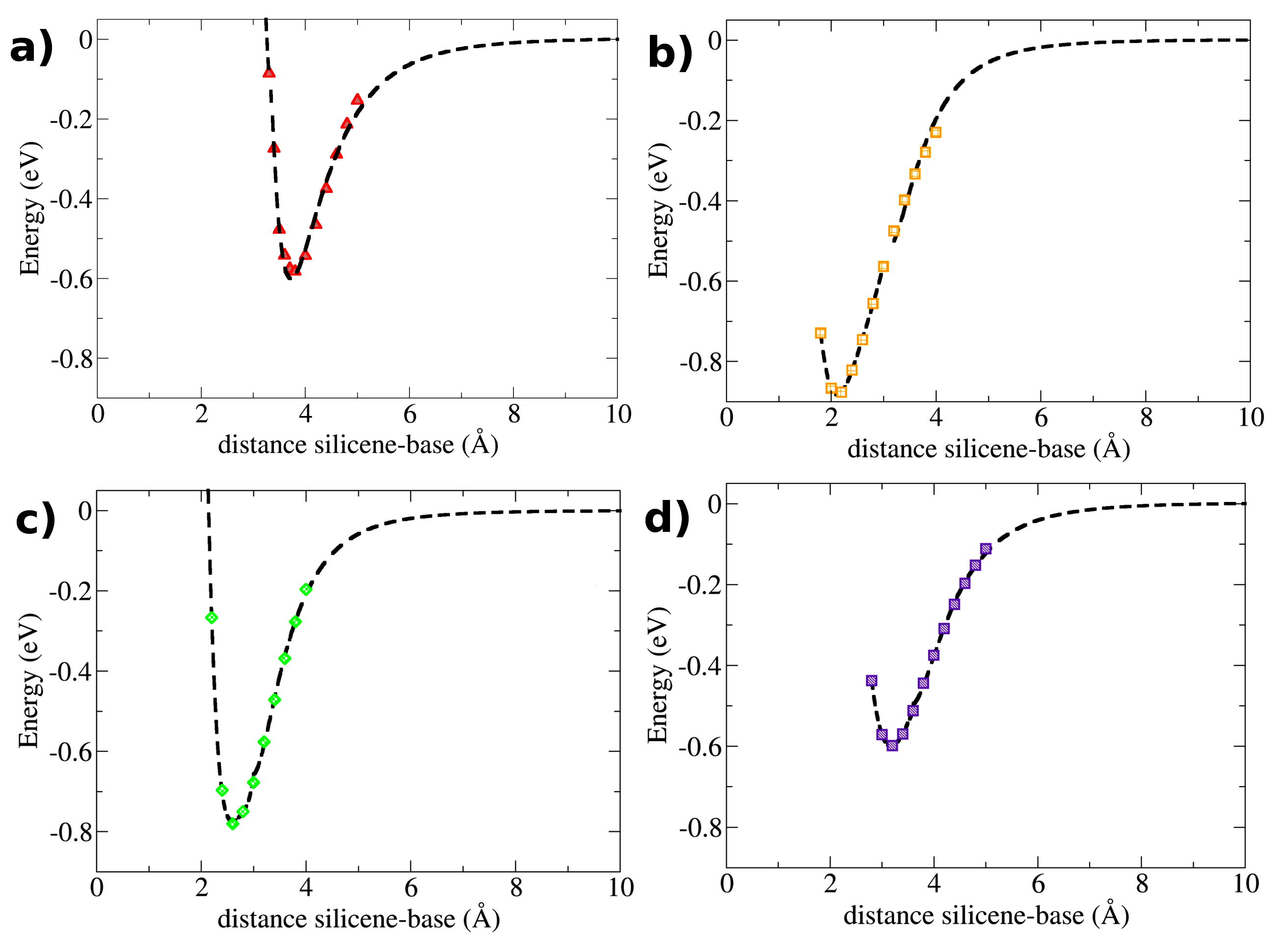}
\caption{Energy variation as a function of the height distance between a nucleobase and the silicene sheet is plotted for each of the four nucleobases: (a) Adenine; (b) Cytosine; (c) Guanine; (d) Thymine. Colored symbols indicate calculated data points; black dashed lines represent a best fit of the data, with a Lennard-Jones type potential for the long-range behavior.}
\label{vertscan}
\end{figure*}

To explore the correlation between equilibrium height of the DNA nucleobases above the silicene substrate and the associated binding energy, we carried out a vertical height scan of the energy of the combined system (Figure \ref{vertscan}). The binding energy was calculated from the difference in energy between the minimum of the fitted curve and the asymptotic limit for large distances. Cytosine is found to exhibit the strongest binding, followed closely by Guanine. Adenine and Thymine are both bound about equally strong with the lowest overall binding energies. The precise numerical values for the binding energies are provided in the last column of Table \ref{tab:distances}. Clearly, the lower binding energies are found for those two nucleobases (A and T) that are furthest away from silicene, as it could be expected.

\section{Charge Density Redistribution}

\begin{figure*}
\hspace{-1cm}
\centering
\includegraphics[scale=0.55]{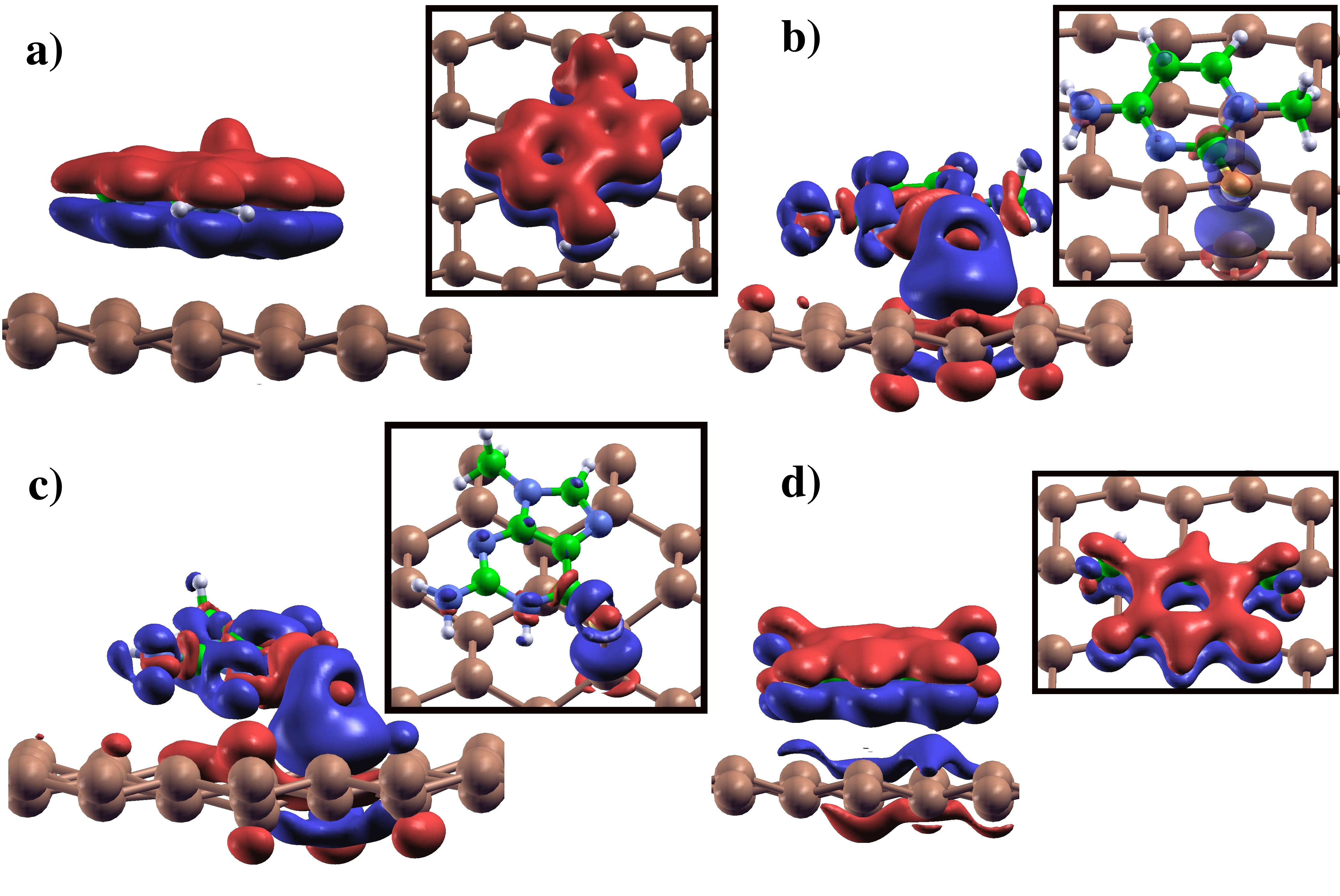}
\caption{The change in electronic charge density is plotted, calculated as the difference between the charge density of the total system (silicene + nucleobase) and that of each constituent part (silicene, nucleobase) separately. Blue color indicates a negative difference in electronic charge density; red color indicates a positive difference in electronic charge density. The main panels show isosurfaces for a value of $0.001$ $electron\times bohr^{-3}$ while the insets show the same charge density difference data plotted for a larger isosurface value of $0.03$ $electron\times bohr^{-3}$ and viewed from a different perspective. (a) Adenine; (b) Cytosine; (c) Guanine; (d) Thymine.}
\label{chargedensity}
\end{figure*}

To better understand the processes occurring at the electronic structure level when a nucleobase interacts with silicene, we calculated how the charge density of the system changes when these two entities are brought together. Mathematically, the change in charge density is simply expressed as $\Delta\rho(\vec{r})=\rho_{Si+base}(\vec{r})-(\rho_{Si}(\vec{r})+\rho_{base}(\vec{r}))$. In Figure \ref{chargedensity} we plot the resulting isosurfaces of the calculated charge density difference in space for the four nucleobases physisorbed on silicene.
Blue color indicates that $\Delta\rho(\vec{r})$ possesses a \emph{negative} value, meaning that electronic charge density has \emph{increased} in this region under the physisorption/chemisorption processes, while red color means that $\Delta\rho(\vec{r})$ is of \emph{positive} value, indicating that the electronic charge density has been \emph{decreased} in this region. It can be seen quite clearly from these plots that for adenine and thymine interacting with silicene  (Figure \ref{chargedensity}a and d), charge redistribution is concentrated mainly to the nucleobase part and in a trivial manner: a shift distributed uniformly over the whole area of the nucleobase, which could be interpreted as an electrostatic repulsion (electronic charge density shifting away from the silicene sheet). The interaction of adenine and thymine with silicene could thus be characterized as weak and purely non-covalent. However, the situation is radically different for the cases of cytosine and guanine interacting with silicene. Here, we notice a complex redistribution of charge density both in the nucleobase part and in the silicene substrate (Figure \ref{chargedensity}b and c). The respective insets in panels b and c of the figure clearly show that the major charge density redistribution occurs along the connection between the single oxygen atom of these two nucleobases and a Si atom of silicene, suggesting the formation of a weak bond of partially covalent nature. In other words: cytosine and guanine appear to be at least weakly chemisorbed on silicene, while adenine and thymine are merely physisorbed (no sign of covalent interaction). It is interesting to note that S. Kilina et al.\cite{Kilina2007} observed similar behavior for nucleobases adsorbed on a Cu(111) surface.

What is common for cytosine and guanine, the two bases that interact more strongly with silicene, is that they each possess one oxygen atom (Figure \ref{bases_on_silicene}). It is via this oxygen atom that they form a quasi-covalent bond with a Si atom in silicene. The two nucleobases which are found by us not to interact strongly with silicene, adenine and thymine, possess either no oxygen atom (adenine) or two oxygen atoms (thymine). In the former case, it is trivially obvious that where there is no oxygen atom, no oxygen-mediated interaction between nucleobase and silicene can take place. And for thymine, forming two covalent bonds simultaneously with Si atoms through its two oxygen atoms is not possible for simple geometric reasons without enforcing any drastic distortions of the bond angles or bond lengths in thymine. 

\section{Effect on Electronic Transport}


We finally come to the question whether silicene may possess any functionality for DNA sequencing. Stated in a more specific way, we are interested in how the electronic transport properties of silicene are affected when the four different nucleobases are physisorbed on its surface. We distinguish two possible transport directions, perpendicular to zigzag rows on the one hand, and perpendicular to armchair rows on the other hand (cf. Fig. \ref{pristine_silicene_energetics}c and \ref{pristine_silicene_energetics}d).

The energy-resolved transmittance $T(E)$ indicates the probability of an electron to be transmitted from one electrode to the other via the scattering region in between. When analyzing in the following the changes in the zero-bias $T(E)$ function upon adsorption of the nucleobases on silicene, it is important to remember the results of the equilibrium binding geometries presented above, which led us to identify two groups of nucleobases in terms of their silicene-nucleobase distance: group I (A and T) with a larger distance between silicene and nucleobase, and group II (C and G) with shorter distances and stronger coupling between the sole oxygen atom from the respective nucleobases and a protruding silicon atom in the buckled silicene sheet.

\begin{figure*}
\hspace{-1cm}
\centering
\includegraphics[scale=0.40]{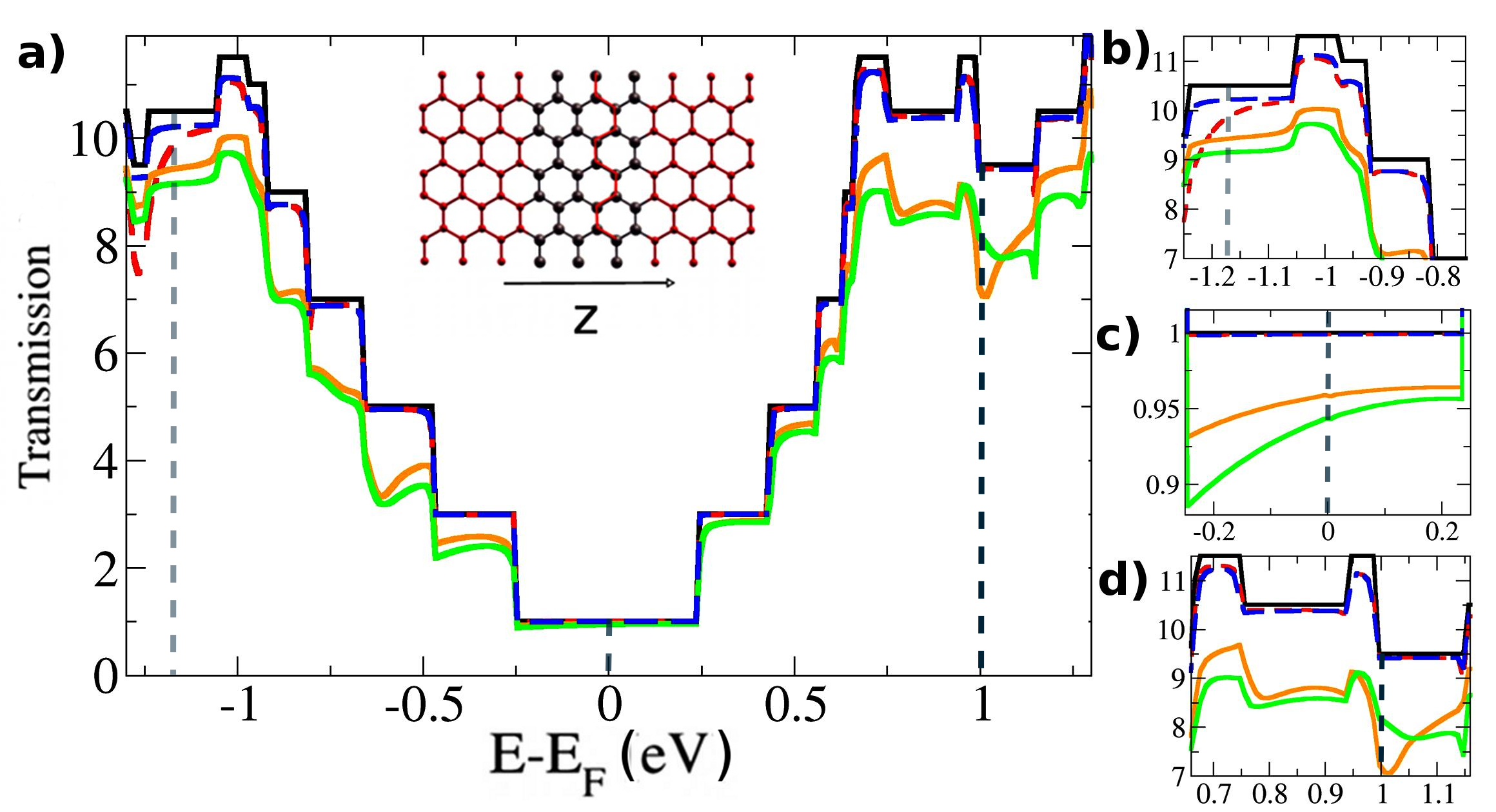}
\caption{The plot shows the zero-bias transmission in the direction perpendicular to armchair rows in silicene as a function of the electron energy, with the Fermi level for the whole system aligned to 0 on the $x$-axis. The curve in black color represents the transmittance for pristine silicene without any nucleobases present while the other colors (red, orange, green, blue) refer respectively to the transmission for each nucleobase (A, C, G, T) physisorbed or chemisorbed on top of silicene. Panel (a) shows the overall data while panels (b), (c), and (d) present a focused view of the same data for selected energy intervals of interest. The significance of the dashed vertical lines is discussed in the main text.}
\label{TransportArmchair}
\end{figure*}

First, we discuss transport perpendicular to armchair rows (see Figure \ref{TransportArmchair}), concentrating on three energies: (i) $E-E_F=-1.16$ eV,
(ii) $E=E_F$ (which corresponds to the zero bias conductance of the system), and (iii) $E-E_F=+1.0$ eV. For (i), we find decreases in the transmittance ranging from 5-15\% for the nucleobases, allowing in principle to distinguish all the nucleobases from each other because the transmittance changes are quite characteristic. We note for (ii) that the transmittance is not changing for group I (A and T) compared with pristine silicene; however in group II (C and G) changes of 5\% (7\%) were seen in the transmittance. In the case for (iii) finally we again observe that it might be possible to distinguish C and G nucleobases, and in this case the decrease in transmission is even larger: about 15\% (27\%), respectively.

\begin{figure*}
\hspace{-1cm}
\centering
\includegraphics[scale=0.45]{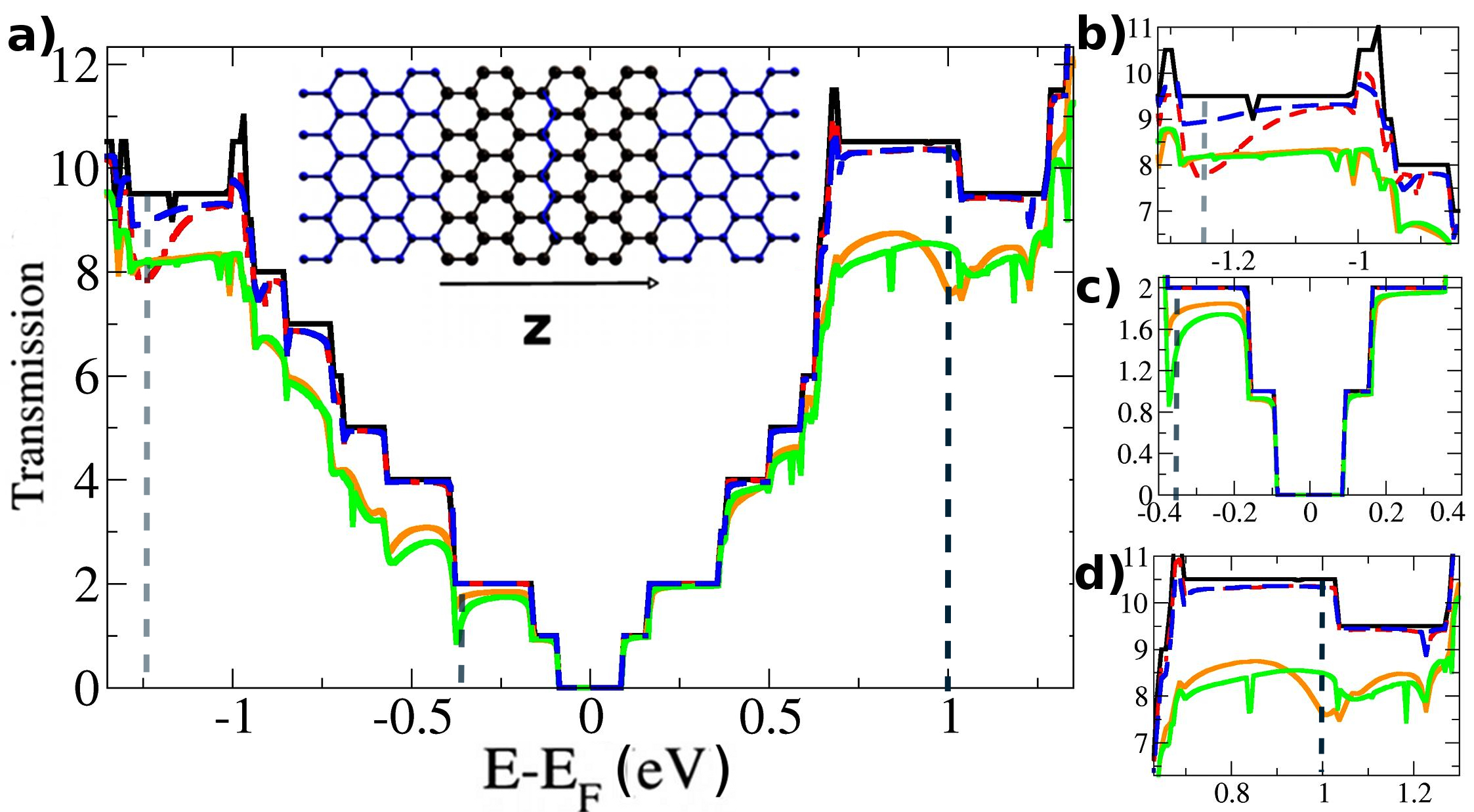}
\caption{The plot shows the zero-bias transmission in the direction perpendicular to zigzag rows in silicene as a function of the electron energy. Otherwise, the same description as in the caption of Figure \ref{TransportArmchair} applies.}
\label{TransportZigzag}
\end{figure*}

Next, we present the analogous analysis for transport perpendicular to zigzag rows (see Figure \ref{TransportZigzag}). Here, we are also concentrating on three energies: (i) $E-E_F=-1.24$ eV and (ii) and (iii) are the same as in the previous case (i.e., $E=E_F$ and $E-E_F=+1.0$ eV). In this case, there is a gap at the Fermi energy (see Figure \ref{TransportZigzag}). The interesting finding here is, for this setup we note that for (i) it is possible to distinguish the A and T nucleobases while for (iii) the other two nucleobases, C and G, show a decrease in the Transmission of 20-25\%.

As expected, group II shows the larger change in transmittance due to the strong coupling with silicene, as seen in the charge density difference plots (cf. Figure \ref{chargedensity}). Group I exhibits a much smaller decrease in the transmittance due to weak interaction with silicene.

\begin{figure*}
\hspace{-1cm}

\includegraphics[scale=0.15]{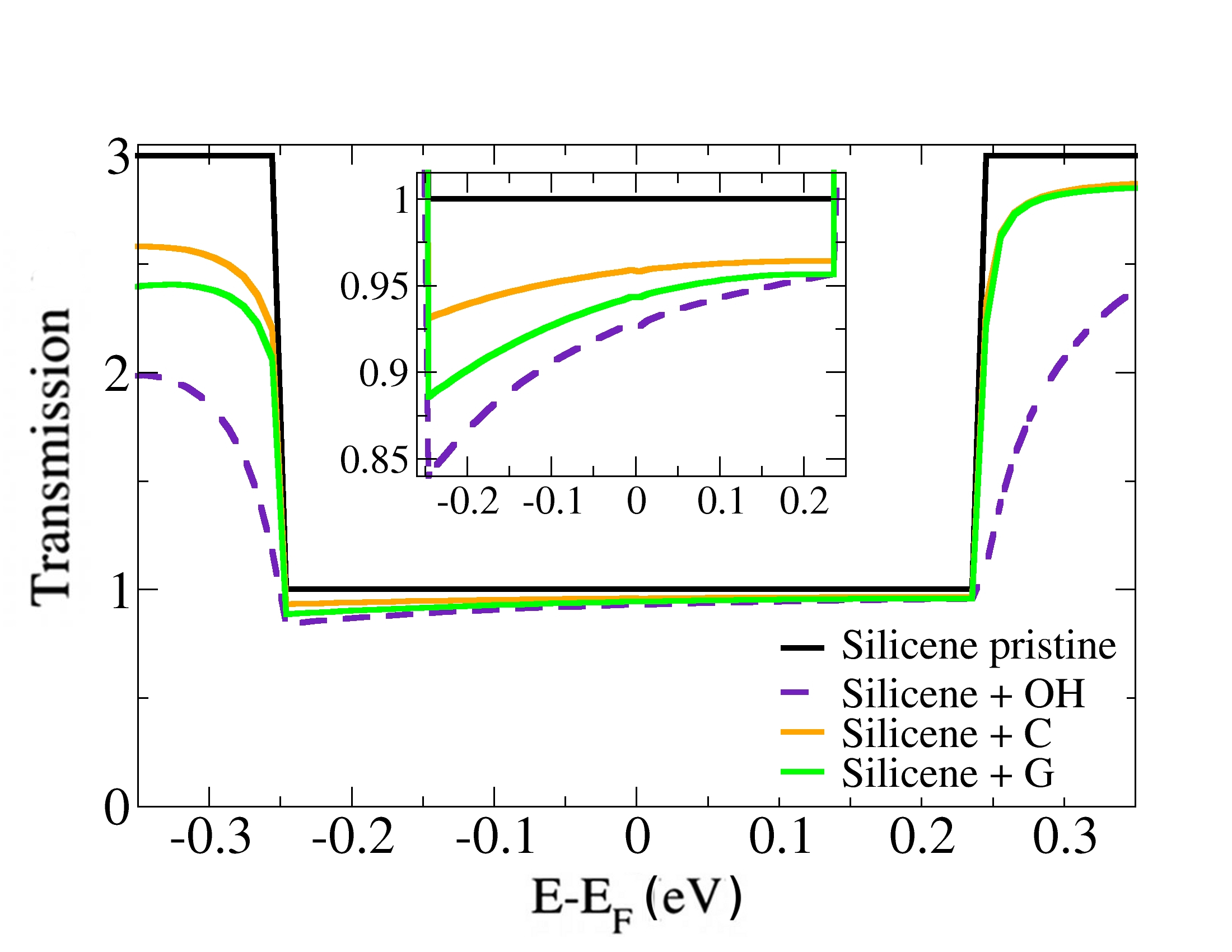}
\caption{Comparison of the transmission as a function of energy between the closely bound nucleobases C and G, and an OH group attached to silicene at the same distance as the oxygen atom in the adsorbed nucleobases C and G.}
\label{transport_OH}
\end{figure*}

To clearly demonstrate that the diminution of the transmittance for C and G is due to the strong coupling between the oxygen atom from the nucleobase and silicene, we performed a transport calculation in a minimalistic model system consisting of an OH group bound with silicene (at the same distance as was found for the system silicene + nucleobase). The transmittance change for silicene when OH is bound to it is shown in Figure \ref{transport_OH} (for comparison, the transmittance plots of C and G have also been reproduced here from Figure \ref{TransportArmchair}). We see that the addition of an OH group to silicene overall generates the same trend as that seen for C and G, thus confirming that the reduction in transmittance originates from the binding of oxygen to silicene.

\section{Conclusions}

In summary, we have presented here a comprehensive theoretical study about DNA nucleobases adsorbed on silicene and discussed a possible emerging application as a biosensor. Our computational results demonstrate that Adenine and Thymine are physisorbed on silicene, whereas Cytosine and Guanine are weakly chemisorbed through the formation of a Si--O bond. Simulated STM images can serve as a guide for experimentalists to recognize nucleobases on silicene. We also investigated the electronic transport using two setups, demonstrating that it is possible to distinguish the four nucleobases. For specific energies, the transmission of the nucleobases differs from that of pristine silicene in a characteristic manner. This result indicates that silicene may be a promising candidate for applications as a biosensor.

\section{Methods}
\label{sec:methodology}

We combined \emph{ab initio} density functional theory (DFT) \cite{Hohenberg1964a,Kohn1965a} as implemented in the SIESTA \cite{Soler2002a} code with the non-equilibrium Green's function (NEGF) method from the TranSiesta code \cite{Brandbyge2002} to perform electronic transport calculations. To take into account weak dispersive interactions we employed a van der Waals correction \cite{Dion2004,Roman-Perez2009} to the Generalized Gradient Approximation (PBE-GGA) \cite{Perdew1996a} for the exchange-correlation functional in DFT. A grid of $6\times1\times4$ $k$-points were used for $k$-space integration and the transport direction is aligned with the $z$-axis. Furthermore, double-$\zeta$ polarized basis sets (DZP) and norm-conserving pseudopotentials \cite{Troullier1991} were used. The conjugate gradient (CG) method was applied to obtain equilibrium structures with residual forces on atoms below $0.01$ eV/\AA.

The principal idea of quantum transport calculations is to divide the system under investigation into three parts: two electrodes and a scattering region in between.
Defining the boundary as a region where the charge density matches with the bulk electrodes and using localized basis sets, it is possible to write the non-equilibrium Green’s functions for the scattering region ${\cal G}\left(E,V\right)$ as:

\begin{equation}
{\cal G}\left(E,V\right) = \left[ E\times S_{\mathrm S} - H_{\mathrm S}\left[\rho\right] - \Sigma_{\mathrm L}\left(E,V\right) - \Sigma_{\mathrm R}\left(E,V\right) \right]^{-1} ~,
\end{equation}

where $S_S$ and $H_S$ are overlap matrix and Hamiltonian, respectively, for the scattering region and $\Sigma_{L/R}$ are self-energies that take into account the effect from the left (L) and right (R) electrode onto the central region. The self energies are given by $\Sigma_{\alpha}=V_{S \alpha}g_{\alpha}V_{\alpha S}$, where $g_{\alpha}$ are the surface Green's functions for the semi-infinite leads and $V_{\alpha S}=V^{\dag}_{S\alpha}$ are the coupling matrix elements between the electrodes and the scatter region. The Hamiltonian can be calculated through a variety of approaches (e.g., using tight-binding methods), but really, $H_S$ is a functional of the electronic density, and for this reason, we used the Hamiltonian obtained from DFT calculations. The charge density is self-consistently calculated via Green's functions until convergence is achieved at which point the transmission coefficient $T(E)$ can be obtained:

\begin{equation}
T\left(E \right) = \Gamma_{\mathrm L}\left(E,V\right) {\cal G}\left(E,V\right) \Gamma_{\mathrm R}\left(E,V\right) {\cal G}^{\dagger}\left(E,V\right)
\end{equation}

where the coupling matrices are given by $\Gamma_\alpha = i \left[ \Sigma_\alpha - \Sigma_\alpha^{\dagger} \right]$, with $\alpha\equiv \left\{{\mathrm{L,R}}\right\}$. Further details regarding the methods for calculating electronic transport properties can be found in the literature \cite{Brandbyge2002,Rocha2006}.

\begin{acknowledgement}
Financial support from the Carl Tryggers Foundation and the Swedish Research Council (VR, Grant No. 621-2009-3628) is gratefully acknowledged. The Swedish National Infrastructure for Computing (SNIC) and the Uppsala Multidisciplinary Center for Advanced Computational Science (UPPMAX) provided computing time for this project.
\end{acknowledgement}

\begin{suppinfo}
Projected Densities of States for the combined system of DNA nucleobases adsorbed in silicene are presented in the supporting information.
\end{suppinfo}

\begin{tocentry}
\includegraphics{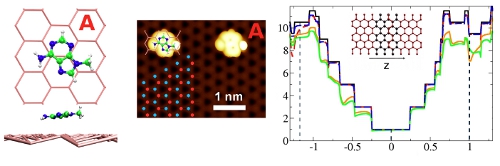}
Structural view and simulated STM image for the nucleobase Adenine physisorbed on silicene. Electronic transport properties for all four DNA nucleobases on silicene are shown to the right.
\end{tocentry}


\bibliography{intro-silicene,methodology,dnaseq,bnnt,addition_references}

\providecommand*\mcitethebibliography{\thebibliography}
\csname @ifundefined\endcsname{endmcitethebibliography}
  {\let\endmcitethebibliography\endthebibliography}{}
\begin{mcitethebibliography}{63}
\providecommand*\natexlab[1]{#1}
\providecommand*\mciteSetBstSublistMode[1]{}
\providecommand*\mciteSetBstMaxWidthForm[2]{}
\providecommand*\mciteBstWouldAddEndPuncttrue
  {\def\EndOfBibitem{\unskip.}}
\providecommand*\mciteBstWouldAddEndPunctfalse
  {\let\EndOfBibitem\relax}
\providecommand*\mciteSetBstMidEndSepPunct[3]{}
\providecommand*\mciteSetBstSublistLabelBeginEnd[3]{}
\providecommand*\EndOfBibitem{}
\mciteSetBstSublistMode{f}
\mciteSetBstMaxWidthForm{subitem}{(\alph{mcitesubitemcount})}
\mciteSetBstSublistLabelBeginEnd
  {\mcitemaxwidthsubitemform\space}
  {\relax}
  {\relax}

\bibitem[Takeda and Shiraishi(1994)Takeda, and Shiraishi]{Takeda1994}
Takeda,~K.; Shiraishi,~K. {Theoretical possibility of stage corrugation in Si
  and Ge analogs of graphite}. \emph{Physical Review B} \textbf{1994},
  \emph{50}, 14916--14922\relax
\mciteBstWouldAddEndPuncttrue
\mciteSetBstMidEndSepPunct{\mcitedefaultmidpunct}
{\mcitedefaultendpunct}{\mcitedefaultseppunct}\relax
\EndOfBibitem
\bibitem[Fagan et~al.(2000)Fagan, Baierle, Mota, da~Silva, and
  Fazzio]{Fagan2000a}
Fagan,~S.; Baierle,~R.; Mota,~R.; da~Silva,~A.; Fazzio,~a. {Ab initio
  calculations for a hypothetical material: Silicon nanotubes}. \emph{Physical
  Review B} \textbf{2000}, \emph{61}, 9994--9996\relax
\mciteBstWouldAddEndPuncttrue
\mciteSetBstMidEndSepPunct{\mcitedefaultmidpunct}
{\mcitedefaultendpunct}{\mcitedefaultseppunct}\relax
\EndOfBibitem
\bibitem[Guzm\'{a}n-Verri and {Lew Yan Voon}(2007)Guzm\'{a}n-Verri, and {Lew
  Yan Voon}]{Guzman-Verri2007a}
Guzm\'{a}n-Verri,~G.; {Lew Yan Voon},~L. {Electronic structure of silicon-based
  nanostructures}. \emph{Physical Review B} \textbf{2007}, \emph{76},
  075131\relax
\mciteBstWouldAddEndPuncttrue
\mciteSetBstMidEndSepPunct{\mcitedefaultmidpunct}
{\mcitedefaultendpunct}{\mcitedefaultseppunct}\relax
\EndOfBibitem
\bibitem[Cahangirov et~al.(2009)Cahangirov, Topsakal, Akt\"{u}rk, \c{S}ahin,
  and Ciraci]{Cahangirov2009}
Cahangirov,~S.; Topsakal,~M.; Akt\"{u}rk,~E.; \c{S}ahin,~H.; Ciraci,~S. {Two-
  and One-Dimensional Honeycomb Structures of Silicon and Germanium}.
  \emph{Physical Review Letters} \textbf{2009}, \emph{102}, 236804\relax
\mciteBstWouldAddEndPuncttrue
\mciteSetBstMidEndSepPunct{\mcitedefaultmidpunct}
{\mcitedefaultendpunct}{\mcitedefaultseppunct}\relax
\EndOfBibitem
\bibitem[Aufray et~al.(2010)Aufray, Kara, Vizzini, Oughaddou, Leandri, Ealet,
  and {Le Lay}]{Aufray2010}
Aufray,~B.; Kara,~A.; Vizzini,~S.; Oughaddou,~H.; Leandri,~C.; Ealet,~B.; {Le
  Lay},~G. {Graphene-like silicon nanoribbons on Ag(110): A possible formation
  of silicene}. \emph{Applied Physics Letters} \textbf{2010}, \emph{96},
  183102\relax
\mciteBstWouldAddEndPuncttrue
\mciteSetBstMidEndSepPunct{\mcitedefaultmidpunct}
{\mcitedefaultendpunct}{\mcitedefaultseppunct}\relax
\EndOfBibitem
\bibitem[Lalmi et~al.(2010)Lalmi, Oughaddou, Enriquez, Kara, Vizzini, Ealet,
  and Aufray]{Lalmi2010a}
Lalmi,~B.; Oughaddou,~H.; Enriquez,~H.; Kara,~A.; Vizzini,~S.; Ealet,~B.;
  Aufray,~B. {Epitaxial growth of a silicene sheet}. \emph{Applied Physics
  Letters} \textbf{2010}, \emph{97}, 223109\relax
\mciteBstWouldAddEndPuncttrue
\mciteSetBstMidEndSepPunct{\mcitedefaultmidpunct}
{\mcitedefaultendpunct}{\mcitedefaultseppunct}\relax
\EndOfBibitem
\bibitem[Lin et~al.(2012)Lin, Arafune, Kawahara, Tsukahara, Minamitani, Kim,
  Takagi, and Kawai]{Lin2012a}
Lin,~C.-L.; Arafune,~R.; Kawahara,~K.; Tsukahara,~N.; Minamitani,~E.; Kim,~Y.;
  Takagi,~N.; Kawai,~M. {Structure of Silicene Grown on Ag(111)}. \emph{Applied
  Physics Express} \textbf{2012}, \emph{5}, 045802\relax
\mciteBstWouldAddEndPuncttrue
\mciteSetBstMidEndSepPunct{\mcitedefaultmidpunct}
{\mcitedefaultendpunct}{\mcitedefaultseppunct}\relax
\EndOfBibitem
\bibitem[Jamgotchian et~al.(2012)Jamgotchian, Colignon, Hamzaoui, Ealet,
  Hoarau, Aufray, and Bib\'{e}rian]{Jamgotchian2012}
Jamgotchian,~H.; Colignon,~Y.; Hamzaoui,~N.; Ealet,~B.; Hoarau,~J.~Y.;
  Aufray,~B.; Bib\'{e}rian,~J.~P. {Growth of silicene layers on Ag(111):
  unexpected effect of the substrate temperature.} \emph{Journal of physics.
  Condensed matter : an Institute of Physics journal} \textbf{2012}, \emph{24},
  172001\relax
\mciteBstWouldAddEndPuncttrue
\mciteSetBstMidEndSepPunct{\mcitedefaultmidpunct}
{\mcitedefaultendpunct}{\mcitedefaultseppunct}\relax
\EndOfBibitem
\bibitem[Feng et~al.(2012)Feng, Ding, Meng, Yao, He, Cheng, Chen, and
  Wu]{Feng2012}
Feng,~B.; Ding,~Z.; Meng,~S.; Yao,~Y.; He,~X.; Cheng,~P.; Chen,~L.; Wu,~K.
  {Evidence of silicene in honeycomb structures of silicon on Ag(111).}
  \emph{Nano letters} \textbf{2012}, \emph{12}, 3507--11\relax
\mciteBstWouldAddEndPuncttrue
\mciteSetBstMidEndSepPunct{\mcitedefaultmidpunct}
{\mcitedefaultendpunct}{\mcitedefaultseppunct}\relax
\EndOfBibitem
\bibitem[Vogt et~al.(2012)Vogt, Padova, Quaresima, Avila, and
  Frantzeskakis]{Vogt2012}
Vogt,~P.; Padova,~P.~D.; Quaresima,~C.; Avila,~J.; Frantzeskakis,~E. {Silicene:
  Compelling Experimental Evidence for Graphenelike Two-Dimensional Silicon}.
  \emph{Physical Review Letters} \textbf{2012}, \emph{108}, 155501\relax
\mciteBstWouldAddEndPuncttrue
\mciteSetBstMidEndSepPunct{\mcitedefaultmidpunct}
{\mcitedefaultendpunct}{\mcitedefaultseppunct}\relax
\EndOfBibitem
\bibitem[Fleurence et~al.(2012)Fleurence, Friedlein, Ozaki, Kawai, Wang, and
  Yamada-Takamura]{Fleurence2012}
Fleurence,~A.; Friedlein,~R.; Ozaki,~T.; Kawai,~H.; Wang,~Y.;
  Yamada-Takamura,~Y. {Experimental Evidence for Epitaxial Silicene on Diboride
  Thin Films}. \emph{Physical Review Letters} \textbf{2012}, \emph{108},
  245501\relax
\mciteBstWouldAddEndPuncttrue
\mciteSetBstMidEndSepPunct{\mcitedefaultmidpunct}
{\mcitedefaultendpunct}{\mcitedefaultseppunct}\relax
\EndOfBibitem
\bibitem[Novoselov et~al.(2004)Novoselov, Geim, Morozov, Jiang, Zhang, Dubonos,
  Grigorieva, and Firsov]{Novoselov2004}
Novoselov,~K.~S.; Geim,~a.~K.; Morozov,~S.~V.; Jiang,~D.; Zhang,~Y.;
  Dubonos,~S.~V.; Grigorieva,~I.~V.; Firsov,~a.~a. {Electric field effect in
  atomically thin carbon films.} \emph{Science (New York, N.Y.)} \textbf{2004},
  \emph{306}, 666--9\relax
\mciteBstWouldAddEndPuncttrue
\mciteSetBstMidEndSepPunct{\mcitedefaultmidpunct}
{\mcitedefaultendpunct}{\mcitedefaultseppunct}\relax
\EndOfBibitem
\bibitem[Branton et~al.(2008)Branton, Deamer, Marziali, Bayley, Benner, Butler,
  {Di Ventra}, Garaj, Hibbs, Huang, Jovanovich, Krstic, Lindsay, Ling,
  Mastrangelo, Meller, Oliver, Pershin, Ramsey, Riehn, Soni, Tabard-Cossa,
  Wanunu, Wiggin, and Schloss]{Branton2008}
Branton,~D.; Deamer,~D.~W.; Marziali,~A.; Bayley,~H.; Benner,~S.~a.;
  Butler,~T.; {Di Ventra},~M.; Garaj,~S.; Hibbs,~A.; Huang,~X. et~al.  {The
  potential and challenges of nanopore sequencing.} \emph{Nature biotechnology}
  \textbf{2008}, \emph{26}, 1146--53\relax
\mciteBstWouldAddEndPuncttrue
\mciteSetBstMidEndSepPunct{\mcitedefaultmidpunct}
{\mcitedefaultendpunct}{\mcitedefaultseppunct}\relax
\EndOfBibitem
\bibitem[Fyta et~al.(2011)Fyta, Melchionna, and Succi]{Fyta2011}
Fyta,~M.; Melchionna,~S.; Succi,~S. {Translocation of biomolecules through
  solid-state nanopores: Theory meets experiments}. \emph{Journal of Polymer
  Science Part B: Polymer Physics} \textbf{2011}, \emph{49}, 985--1011\relax
\mciteBstWouldAddEndPuncttrue
\mciteSetBstMidEndSepPunct{\mcitedefaultmidpunct}
{\mcitedefaultendpunct}{\mcitedefaultseppunct}\relax
\EndOfBibitem
\bibitem[Venkatesan and Bashir(2011)Venkatesan, and Bashir]{Venkatesan2011}
Venkatesan,~B.~M.; Bashir,~R. {Nanopore sensors for nucleic acid analysis.}
  \emph{Nature nanotechnology} \textbf{2011}, \emph{6}, 615--24\relax
\mciteBstWouldAddEndPuncttrue
\mciteSetBstMidEndSepPunct{\mcitedefaultmidpunct}
{\mcitedefaultendpunct}{\mcitedefaultseppunct}\relax
\EndOfBibitem
\bibitem[Scheicher et~al.({2012})Scheicher, Grigoriev, and
  Ahuja]{Scheicher2012}
Scheicher,~R.~H.; Grigoriev,~A.; Ahuja,~R. {DNA sequencing with nanopores from
  an ab initio perspective}. \emph{{JOURNAL OF MATERIALS SCIENCE}}
  \textbf{{2012}}, \emph{{47}}, {7439--7446}\relax
\mciteBstWouldAddEndPuncttrue
\mciteSetBstMidEndSepPunct{\mcitedefaultmidpunct}
{\mcitedefaultendpunct}{\mcitedefaultseppunct}\relax
\EndOfBibitem
\bibitem[{Di Ventra}(2013)]{DiVentra2013}
{Di Ventra},~M. {Fast DNA sequencing by electrical means inches closer.}
  \emph{Nanotechnology} \textbf{2013}, \emph{24}, 342501\relax
\mciteBstWouldAddEndPuncttrue
\mciteSetBstMidEndSepPunct{\mcitedefaultmidpunct}
{\mcitedefaultendpunct}{\mcitedefaultseppunct}\relax
\EndOfBibitem
\bibitem[Yang et~al.({2013})Yang, Liu, Xie, Hui, Jiao, Gong, and
  Zhang]{Yang2013}
Yang,~Y.; Liu,~R.; Xie,~H.; Hui,~Y.; Jiao,~R.; Gong,~Y.; Zhang,~Y. {Advances in
  Nanopore Sequencing Technology}. \emph{{JOURNAL OF NANOSCIENCE AND
  NANOTECHNOLOGY}} \textbf{{2013}}, \emph{{13}}, {4521--4538}\relax
\mciteBstWouldAddEndPuncttrue
\mciteSetBstMidEndSepPunct{\mcitedefaultmidpunct}
{\mcitedefaultendpunct}{\mcitedefaultseppunct}\relax
\EndOfBibitem
\bibitem[Haque et~al.({2013})Haque, Li, Wu, Liang, and Guo]{Haque2013}
Haque,~F.; Li,~J.; Wu,~H.-C.; Liang,~X.-J.; Guo,~P. {Solid-state and biological
  nanopore for real-time sensing of single chemical and sequencing of DNA}.
  \emph{{NANO TODAY}} \textbf{{2013}}, \emph{{8}}, {56--74}\relax
\mciteBstWouldAddEndPuncttrue
\mciteSetBstMidEndSepPunct{\mcitedefaultmidpunct}
{\mcitedefaultendpunct}{\mcitedefaultseppunct}\relax
\EndOfBibitem
\bibitem[Postma(2010)]{Postma2010}
Postma,~H. W.~C. {Rapid sequencing of individual DNA molecules in graphene
  nanogaps.} \emph{Nano letters} \textbf{2010}, \emph{10}, 420--5\relax
\mciteBstWouldAddEndPuncttrue
\mciteSetBstMidEndSepPunct{\mcitedefaultmidpunct}
{\mcitedefaultendpunct}{\mcitedefaultseppunct}\relax
\EndOfBibitem
\bibitem[Merchant et~al.(2010)Merchant, Healy, Wanunu, Ray, Peterman, Bartel,
  Fischbein, Venta, Luo, Johnson, and Drndi\'{c}]{Merchant2010}
Merchant,~C.~a.; Healy,~K.; Wanunu,~M.; Ray,~V.; Peterman,~N.; Bartel,~J.;
  Fischbein,~M.~D.; Venta,~K.; Luo,~Z.; Johnson,~a. T.~C. et~al.  {DNA
  translocation through graphene nanopores.} \emph{Nano letters} \textbf{2010},
  \emph{10}, 2915--21\relax
\mciteBstWouldAddEndPuncttrue
\mciteSetBstMidEndSepPunct{\mcitedefaultmidpunct}
{\mcitedefaultendpunct}{\mcitedefaultseppunct}\relax
\EndOfBibitem
\bibitem[Schneider et~al.(2010)Schneider, Kowalczyk, Calado, Pandraud,
  Zandbergen, Vandersypen, and Dekker]{Schneider2010}
Schneider,~G.~F.; Kowalczyk,~S.~W.; Calado,~V.~E.; Pandraud,~G.;
  Zandbergen,~H.~W.; Vandersypen,~L. M.~K.; Dekker,~C. {DNA translocation
  through graphene nanopores.} \emph{Nano letters} \textbf{2010}, \emph{10},
  3163--7\relax
\mciteBstWouldAddEndPuncttrue
\mciteSetBstMidEndSepPunct{\mcitedefaultmidpunct}
{\mcitedefaultendpunct}{\mcitedefaultseppunct}\relax
\EndOfBibitem
\bibitem[Garaj et~al.(2010)Garaj, Hubbard, Reina, Kong, Branton, and
  Golovchenko]{Garaj2010}
Garaj,~S.; Hubbard,~W.; Reina,~a.; Kong,~J.; Branton,~D.; Golovchenko,~J.~a.
  {Graphene as a subnanometre trans-electrode membrane.} \emph{Nature}
  \textbf{2010}, \emph{467}, 190--3\relax
\mciteBstWouldAddEndPuncttrue
\mciteSetBstMidEndSepPunct{\mcitedefaultmidpunct}
{\mcitedefaultendpunct}{\mcitedefaultseppunct}\relax
\EndOfBibitem
\bibitem[Nelson et~al.(2010)Nelson, Zhang, and Prezhdo]{Nelson2010}
Nelson,~T.; Zhang,~B.; Prezhdo,~O.~V. {Detection of nucleic acids with graphene
  nanopores: ab initio characterization of a novel sequencing device.}
  \emph{Nano letters} \textbf{2010}, \emph{10}, 3237--42\relax
\mciteBstWouldAddEndPuncttrue
\mciteSetBstMidEndSepPunct{\mcitedefaultmidpunct}
{\mcitedefaultendpunct}{\mcitedefaultseppunct}\relax
\EndOfBibitem
\bibitem[Min et~al.(2011)Min, Kim, Cho, and Kim]{Min2011}
Min,~S.~K.; Kim,~W.~Y.; Cho,~Y.; Kim,~K.~S. {Fast DNA sequencing with a
  graphene-based nanochannel device.} \emph{Nature nanotechnology}
  \textbf{2011}, \emph{6}, 162--5\relax
\mciteBstWouldAddEndPuncttrue
\mciteSetBstMidEndSepPunct{\mcitedefaultmidpunct}
{\mcitedefaultendpunct}{\mcitedefaultseppunct}\relax
\EndOfBibitem
\bibitem[Cho et~al.(2011)Cho, Min, Kim, and Kim]{Cho2011}
Cho,~Y.; Min,~S.~K.; Kim,~W.~Y.; Kim,~K.~S. {The origin of dips for the
  graphene-based DNA sequencing device.} \emph{Physical chemistry chemical
  physics : PCCP} \textbf{2011}, \emph{13}, 14293--6\relax
\mciteBstWouldAddEndPuncttrue
\mciteSetBstMidEndSepPunct{\mcitedefaultmidpunct}
{\mcitedefaultendpunct}{\mcitedefaultseppunct}\relax
\EndOfBibitem
\bibitem[Bergvall et~al.({2011})Bergvall, Berland, Hyldgaard, Kubatkin, and
  Lofwander]{Bergvall2011}
Bergvall,~A.; Berland,~K.; Hyldgaard,~P.; Kubatkin,~S.; Lofwander,~T. {Graphene
  nanogap for gate-tunable quantum-coherent single-molecule electronics}.
  \emph{{PHYSICAL REVIEW B}} \textbf{{2011}}, \emph{{84}}, {155451}\relax
\mciteBstWouldAddEndPuncttrue
\mciteSetBstMidEndSepPunct{\mcitedefaultmidpunct}
{\mcitedefaultendpunct}{\mcitedefaultseppunct}\relax
\EndOfBibitem
\bibitem[Sathe et~al.({2011})Sathe, Zou, Leburton, and Schulten]{Sathe2011}
Sathe,~C.; Zou,~X.; Leburton,~J.-P.; Schulten,~K. {Computational Investigation
  of DNA Detection Using Graphene Nanopores}. \emph{{ACS NANO}}
  \textbf{{2011}}, \emph{{5}}, {8842--8851}\relax
\mciteBstWouldAddEndPuncttrue
\mciteSetBstMidEndSepPunct{\mcitedefaultmidpunct}
{\mcitedefaultendpunct}{\mcitedefaultseppunct}\relax
\EndOfBibitem
\bibitem[He et~al.(2011)He, Scheicher, Grigoriev, Ahuja, Long, Huo, and
  Liu]{He2011}
He,~Y.; Scheicher,~R.~H.; Grigoriev,~A.; Ahuja,~R.; Long,~S.; Huo,~Z.; Liu,~M.
  {Enhanced DNA Sequencing Performance Through Edge-Hydrogenation of Graphene
  Electrodes}. \emph{Advanced Functional Materials} \textbf{2011}, \emph{21},
  2674--2679\relax
\mciteBstWouldAddEndPuncttrue
\mciteSetBstMidEndSepPunct{\mcitedefaultmidpunct}
{\mcitedefaultendpunct}{\mcitedefaultseppunct}\relax
\EndOfBibitem
\bibitem[Prasongkit et~al.(2011)Prasongkit, Grigoriev, Pathak, Ahuja, and
  Scheicher]{Prasongkit2011}
Prasongkit,~J.; Grigoriev,~A.; Pathak,~B.; Ahuja,~R.; Scheicher,~R.~H.
  {Transverse conductance of DNA nucleotides in a graphene nanogap from first
  principles.} \emph{Nano Letters} \textbf{2011}, \emph{11}, 1941--5\relax
\mciteBstWouldAddEndPuncttrue
\mciteSetBstMidEndSepPunct{\mcitedefaultmidpunct}
{\mcitedefaultendpunct}{\mcitedefaultseppunct}\relax
\EndOfBibitem
\bibitem[Liu et~al.(2012)Liu, Dong, and Chen]{Liu2012a}
Liu,~Y.; Dong,~X.; Chen,~P. {Biological and chemical sensors based on graphene
  materials.} \emph{Chemical Society reviews} \textbf{2012}, \emph{41},
  2283--307\relax
\mciteBstWouldAddEndPuncttrue
\mciteSetBstMidEndSepPunct{\mcitedefaultmidpunct}
{\mcitedefaultendpunct}{\mcitedefaultseppunct}\relax
\EndOfBibitem
\bibitem[Saha et~al.(2012)Saha, Drndi\'{c}, and Nikoli\'{c}]{Saha2012}
Saha,~K.~K.; Drndi\'{c},~M.; Nikoli\'{c},~B.~K. {DNA base-specific modulation
  of microampere transverse edge currents through a metallic graphene
  nanoribbon with a nanopore.} \emph{Nano letters} \textbf{2012}, \emph{12},
  50--5\relax
\mciteBstWouldAddEndPuncttrue
\mciteSetBstMidEndSepPunct{\mcitedefaultmidpunct}
{\mcitedefaultendpunct}{\mcitedefaultseppunct}\relax
\EndOfBibitem
\bibitem[Venkatesan et~al.({2012})Venkatesan, Estrada, Banerjee, Jin, Dorgan,
  Bae, Aluru, Pop, and Bashir]{Venkatesan2012}
Venkatesan,~B.~M.; Estrada,~D.; Banerjee,~S.; Jin,~X.; Dorgan,~V.~E.;
  Bae,~M.-H.; Aluru,~N.~R.; Pop,~E.; Bashir,~R. {Stacked Graphene-Al2O3
  Nanopore Sensors for Sensitive Detection of DNA and DNA-Protein Complexes}.
  \emph{{ACS NANO}} \textbf{{2012}}, \emph{{6}}, {441--450}\relax
\mciteBstWouldAddEndPuncttrue
\mciteSetBstMidEndSepPunct{\mcitedefaultmidpunct}
{\mcitedefaultendpunct}{\mcitedefaultseppunct}\relax
\EndOfBibitem
\bibitem[Cheng and Zhao({2012})Cheng, and Zhao]{Cheng2012}
Cheng,~C.-L.; Zhao,~G.-J. {Steered molecular dynamics simulation study on
  dynamic self-assembly of single-stranded DNA with double-walled carbon
  nanotube and graphene}. \emph{{NANOSCALE}} \textbf{{2012}}, \emph{{4}},
  {2301--2305}\relax
\mciteBstWouldAddEndPuncttrue
\mciteSetBstMidEndSepPunct{\mcitedefaultmidpunct}
{\mcitedefaultendpunct}{\mcitedefaultseppunct}\relax
\EndOfBibitem
\bibitem[Qiu and Guo({2012})Qiu, and Guo]{Qiu2012}
Qiu,~H.; Guo,~W. {Detecting ssDNA at single-nucleotide resolution by
  sub-2-nanometer pore in monoatomic graphene: A molecular dynamics study}.
  \emph{{APPLIED PHYSICS LETTERS}} \textbf{{2012}}, \emph{{100}},
  {083106}\relax
\mciteBstWouldAddEndPuncttrue
\mciteSetBstMidEndSepPunct{\mcitedefaultmidpunct}
{\mcitedefaultendpunct}{\mcitedefaultseppunct}\relax
\EndOfBibitem
\bibitem[Wells et~al.({2012})Wells, Belkin, Comer, and Aksimentiev]{Wells2012}
Wells,~D.~B.; Belkin,~M.; Comer,~J.; Aksimentiev,~A. {Assessing Graphene
  Nanopores for Sequencing DNA}. \emph{{NANO LETTERS}} \textbf{{2012}},
  \emph{{12}}, {4117--4123}\relax
\mciteBstWouldAddEndPuncttrue
\mciteSetBstMidEndSepPunct{\mcitedefaultmidpunct}
{\mcitedefaultendpunct}{\mcitedefaultseppunct}\relax
\EndOfBibitem
\bibitem[Liu et~al.({2012})Liu, Dong, and Chen]{Liu2012}
Liu,~Y.; Dong,~X.; Chen,~P. {Biological and chemical sensors based on graphene
  materials}. \emph{{CHEMICAL SOCIETY REVIEWS}} \textbf{{2012}}, \emph{{41}},
  {2283--2307}\relax
\mciteBstWouldAddEndPuncttrue
\mciteSetBstMidEndSepPunct{\mcitedefaultmidpunct}
{\mcitedefaultendpunct}{\mcitedefaultseppunct}\relax
\EndOfBibitem
\bibitem[Lv et~al.({2013})Lv, Chen, and Wu]{Lv2013}
Lv,~W.; Chen,~M.; Wu,~R. {The impact of the number of layers of a graphene
  nanopore on DNA translocation}. \emph{{SOFT MATTER}} \textbf{{2013}},
  \emph{{9}}, {960--966}\relax
\mciteBstWouldAddEndPuncttrue
\mciteSetBstMidEndSepPunct{\mcitedefaultmidpunct}
{\mcitedefaultendpunct}{\mcitedefaultseppunct}\relax
\EndOfBibitem
\bibitem[Garaj et~al.({2013})Garaj, Liu, Golovchenko, and
  Branton]{SlavenGaraj2013}
Garaj,~S.; Liu,~S.; Golovchenko,~J.~A.; Branton,~D. {Molecule-hugging graphene
  nanopores}. \emph{{Proceedings of the National Academy of Sciences of tThe
  United States of America}} \textbf{{2013}}, \emph{{110}},
  {12192--12196}\relax
\mciteBstWouldAddEndPuncttrue
\mciteSetBstMidEndSepPunct{\mcitedefaultmidpunct}
{\mcitedefaultendpunct}{\mcitedefaultseppunct}\relax
\EndOfBibitem
\bibitem[Li et~al.({2013})Li, Zhang, Yang, Bi, Ni, Li, and Chen]{Li2013}
Li,~J.; Zhang,~Y.; Yang,~J.; Bi,~K.; Ni,~Z.; Li,~D.; Chen,~Y. {Molecular
  dynamics study of DNA translocation through graphene nanopores}.
  \emph{{PHYSICAL REVIEW E}} \textbf{{2013}}, \emph{{87}}, {062707}\relax
\mciteBstWouldAddEndPuncttrue
\mciteSetBstMidEndSepPunct{\mcitedefaultmidpunct}
{\mcitedefaultendpunct}{\mcitedefaultseppunct}\relax
\EndOfBibitem
\bibitem[Hui et~al.({2013})Hui, Xiang, Sheng-Lin, Jun, Yu, and
  Fang-Ping]{Hui2013}
Hui,~Z.; Xiang,~N.; Sheng-Lin,~P.; Jun,~O.; Yu,~C.; Fang-Ping,~O.
  {First-Principles Study of Graphene-Based Biomolecular Sensor}. \emph{{ACTA
  PHYSICO-CHIMICA SINICA}} \textbf{{2013}}, \emph{{29}}, {250--254}\relax
\mciteBstWouldAddEndPuncttrue
\mciteSetBstMidEndSepPunct{\mcitedefaultmidpunct}
{\mcitedefaultendpunct}{\mcitedefaultseppunct}\relax
\EndOfBibitem
\bibitem[Avdoshenko et~al.({2013})Avdoshenko, Nozaki, da~Rocha, Gonzalez, Lee,
  Gutierrez, and Cuniberti]{Avdoshenko2013}
Avdoshenko,~S.~M.; Nozaki,~D.; da~Rocha,~C.~G.; Gonzalez,~J.~W.; Lee,~M.~H.;
  Gutierrez,~R.; Cuniberti,~G. {Dynamic and Electronic Transport Properties of
  DNA Translocation through Graphene Nanopores}. \emph{{NANO LETTERS}}
  \textbf{{2013}}, \emph{{13}}, {1969--1976}\relax
\mciteBstWouldAddEndPuncttrue
\mciteSetBstMidEndSepPunct{\mcitedefaultmidpunct}
{\mcitedefaultendpunct}{\mcitedefaultseppunct}\relax
\EndOfBibitem
\bibitem[Prasongkit et~al.({2013})Prasongkit, Grigoriev, Pathak, Ahuja, and
  Scheicher]{Prasongkit2013}
Prasongkit,~J.; Grigoriev,~A.; Pathak,~B.; Ahuja,~R.; Scheicher,~R.~H.
  {Theoretical Study of Electronic Transport through DNA Nucleotides in a
  Double-Functionalized Graphene Nanogap}. \emph{{JOURNAL OF PHYSICAL CHEMISTRY
  C}} \textbf{{2013}}, \emph{{117}}, {15421--15428}\relax
\mciteBstWouldAddEndPuncttrue
\mciteSetBstMidEndSepPunct{\mcitedefaultmidpunct}
{\mcitedefaultendpunct}{\mcitedefaultseppunct}\relax
\EndOfBibitem
\bibitem[Jeong et~al.({2013})Jeong, Kim, Lee, Lee, Kim, and Huh]{Jeong2013}
Jeong,~H.; Kim,~H.~S.; Lee,~S.-H.; Lee,~D.; Kim,~Y.~H.; Huh,~N. {Quantum
  interference in DNA bases probed by graphene nanoribbons}. \emph{{APPLIED
  PHYSICS LETTERS}} \textbf{{2013}}, \emph{{103}}, {023701}\relax
\mciteBstWouldAddEndPuncttrue
\mciteSetBstMidEndSepPunct{\mcitedefaultmidpunct}
{\mcitedefaultendpunct}{\mcitedefaultseppunct}\relax
\EndOfBibitem
\bibitem[Freedman et~al.({2013})Freedman, Ahn, and Kim]{Freedman2013}
Freedman,~K.~J.; Ahn,~C.~W.; Kim,~M.~J. {Detection of Long and Short DNA Using
  Nanopores with Graphitic Polyhedral Edges}. \emph{{ACS NANO}}
  \textbf{{2013}}, \emph{{7}}, {5008--5016}\relax
\mciteBstWouldAddEndPuncttrue
\mciteSetBstMidEndSepPunct{\mcitedefaultmidpunct}
{\mcitedefaultendpunct}{\mcitedefaultseppunct}\relax
\EndOfBibitem
\bibitem[Schneider et~al.({2013})Schneider, Xu, Hage, Luik, Spoor, Malladi,
  Zandbergen, and Dekker]{Dekker2013}
Schneider,~G.~F.; Xu,~Q.; Hage,~S.; Luik,~S.; Spoor,~J. N.~H.; Malladi,~S.;
  Zandbergen,~H.; Dekker,~C. {Tailoring the hydrophobicity of graphene for its
  use as nanopores for DNA translocation}. \emph{{NATURE COMMUNICATIONS}}
  \textbf{{2013}}, \emph{{4}}, {2619}\relax
\mciteBstWouldAddEndPuncttrue
\mciteSetBstMidEndSepPunct{\mcitedefaultmidpunct}
{\mcitedefaultendpunct}{\mcitedefaultseppunct}\relax
\EndOfBibitem
\bibitem[Traversi et~al.(2013)Traversi, Raillon, Benameur, Liu, Khlybov, Tosun,
  Krasnozhon, Kis, and Radenovic]{Traversi2013}
Traversi,~F.; Raillon,~C.; Benameur,~S.~M.; Liu,~K.; Khlybov,~S.; Tosun,~M.;
  Krasnozhon,~D.; Kis,~A.; Radenovic,~A. {Detecting the translocation of DNA
  through a nanopore using graphene nanoribbons}. \emph{Nature Nanotechnology}
  \textbf{2013}, \emph{8}, 939--945\relax
\mciteBstWouldAddEndPuncttrue
\mciteSetBstMidEndSepPunct{\mcitedefaultmidpunct}
{\mcitedefaultendpunct}{\mcitedefaultseppunct}\relax
\EndOfBibitem
\bibitem[Mukhopadhyay et~al.(2010)Mukhopadhyay, Gowtham, Scheicher, Pandey, and
  Karna]{Mukhopadhyay2010}
Mukhopadhyay,~S.; Gowtham,~S.; Scheicher,~R.~H.; Pandey,~R.; Karna,~S.~P.
  {Theoretical study of physisorption of nucleobases on boron nitride
  nanotubes: a new class of hybrid nano-biomaterials.} \emph{Nanotechnology}
  \textbf{2010}, \emph{21}, 165703\relax
\mciteBstWouldAddEndPuncttrue
\mciteSetBstMidEndSepPunct{\mcitedefaultmidpunct}
{\mcitedefaultendpunct}{\mcitedefaultseppunct}\relax
\EndOfBibitem
\bibitem[Lee et~al.(2013)Lee, Choi, Kim, Scheicher, and Cho]{Lee2013}
Lee,~J.-H.; Choi,~Y.-K.; Kim,~H.-J.; Scheicher,~R.~H.; Cho,~J.-H.
  {Physisorption of DNA Nucleobases on h -BN and Graphene: vdW-Corrected DFT
  Calculations}. \emph{The Journal of Physical Chemistry C} \textbf{2013},
  \emph{117}, 13435--13441\relax
\mciteBstWouldAddEndPuncttrue
\mciteSetBstMidEndSepPunct{\mcitedefaultmidpunct}
{\mcitedefaultendpunct}{\mcitedefaultseppunct}\relax
\EndOfBibitem
\bibitem[Gowtham et~al.(2007)Gowtham, Scheicher, Ahuja, Pandey, and
  Karna]{Gowtham2007a}
Gowtham,~S.; Scheicher,~R.; Ahuja,~R.; Pandey,~R.; Karna,~S. {Physisorption of
  nucleobases on graphene: Density-functional calculations}. \emph{Physical
  Review B} \textbf{2007}, \emph{76}, 033401\relax
\mciteBstWouldAddEndPuncttrue
\mciteSetBstMidEndSepPunct{\mcitedefaultmidpunct}
{\mcitedefaultendpunct}{\mcitedefaultseppunct}\relax
\EndOfBibitem
\bibitem[Horcas et~al.(2007)Horcas, Fern\'{a}ndez, G\'{o}mez-Rodr\'{\i}guez,
  Colchero, G\'{o}mez-Herrero, and Baro]{Horcas2007}
Horcas,~I.; Fern\'{a}ndez,~R.; G\'{o}mez-Rodr\'{\i}guez,~J.~M.; Colchero,~J.;
  G\'{o}mez-Herrero,~J.; Baro,~a.~M. {WSXM: a software for scanning probe
  microscopy and a tool for nanotechnology.} \emph{The Review of scientific
  instruments} \textbf{2007}, \emph{78}, 013705\relax
\mciteBstWouldAddEndPuncttrue
\mciteSetBstMidEndSepPunct{\mcitedefaultmidpunct}
{\mcitedefaultendpunct}{\mcitedefaultseppunct}\relax
\EndOfBibitem
\bibitem[{Tersoff, J and Hamann}(1985)]{TersoffJandHamann1985}
{Tersoff, J and Hamann},~D.~R. {Theory of the scanning tunneling microscope}.
  \emph{Phys. Rev. B} \textbf{1985}, \emph{31}, 805\relax
\mciteBstWouldAddEndPuncttrue
\mciteSetBstMidEndSepPunct{\mcitedefaultmidpunct}
{\mcitedefaultendpunct}{\mcitedefaultseppunct}\relax
\EndOfBibitem
\bibitem[Kilina et~al.(2007)Kilina, Tretiak, Yarotski, Zhu, Modine, Taylor, and
  a.V. Balatsky]{Kilina2007}
Kilina,~S.; Tretiak,~S.; Yarotski,~D.; Zhu,~J.-X.; Modine,~N.; Taylor,~a.; a.V.
  Balatsky, {Electronic Properties of DNA Base Molecules Adsorbed on a Metallic
  Surface}. \emph{Journal of Physical Chemistry C} \textbf{2007}, \emph{111},
  14541--14551\relax
\mciteBstWouldAddEndPuncttrue
\mciteSetBstMidEndSepPunct{\mcitedefaultmidpunct}
{\mcitedefaultendpunct}{\mcitedefaultseppunct}\relax
\EndOfBibitem
\bibitem[Hohenberg and Kohn(1964)Hohenberg, and Kohn]{Hohenberg1964a}
Hohenberg,~P.; Kohn,~W. {Inhomogeneous electron gas}. \emph{Physical Review}
  \textbf{1964}, \emph{155}\relax
\mciteBstWouldAddEndPuncttrue
\mciteSetBstMidEndSepPunct{\mcitedefaultmidpunct}
{\mcitedefaultendpunct}{\mcitedefaultseppunct}\relax
\EndOfBibitem
\bibitem[Kohn and Sham(1965)Kohn, and Sham]{Kohn1965a}
Kohn,~W.; Sham,~L. {Self-consistent equations including exchange and
  correlation effects}. \emph{PHYSICAL REVIEW} \textbf{1965}, \emph{385}\relax
\mciteBstWouldAddEndPuncttrue
\mciteSetBstMidEndSepPunct{\mcitedefaultmidpunct}
{\mcitedefaultendpunct}{\mcitedefaultseppunct}\relax
\EndOfBibitem
\bibitem[Soler et~al.(2002)Soler, Artacho, Gale, Garc, Junquera, Ordej, and
  Daniel]{Soler2002a}
Soler,~M.; Artacho,~E.; Gale,~J.~D.; Garc,~A.; Junquera,~J.; Ordej,~P.;
  Daniel,~S. {The SIESTA method for ab initio order- N materials}. \emph{J.
  Phys.: Condens. Matter} \textbf{2002}, \emph{2745}, 2745--2779\relax
\mciteBstWouldAddEndPuncttrue
\mciteSetBstMidEndSepPunct{\mcitedefaultmidpunct}
{\mcitedefaultendpunct}{\mcitedefaultseppunct}\relax
\EndOfBibitem
\bibitem[Brandbyge et~al.(2002)Brandbyge, Mozos, Ordej\'{o}n, Taylor, and
  Stokbro]{Brandbyge2002}
Brandbyge,~M.; Mozos,~J.-L.; Ordej\'{o}n,~P.; Taylor,~J.; Stokbro,~K.
  {Density-functional method for nonequilibrium electron transport}.
  \emph{Physical Review B} \textbf{2002}, \emph{65}, 165401\relax
\mciteBstWouldAddEndPuncttrue
\mciteSetBstMidEndSepPunct{\mcitedefaultmidpunct}
{\mcitedefaultendpunct}{\mcitedefaultseppunct}\relax
\EndOfBibitem
\bibitem[Dion et~al.(2004)Dion, Rydberg, Schr\"{o}der, Langreth, and
  Lundqvist]{Dion2004}
Dion,~M.; Rydberg,~H.; Schr\"{o}der,~E.; Langreth,~D.~C.; Lundqvist,~B.~I. {Van
  der Waals Density Functional for General Geometries}. \emph{Physical Review
  Letters} \textbf{2004}, \emph{92}, 246401\relax
\mciteBstWouldAddEndPuncttrue
\mciteSetBstMidEndSepPunct{\mcitedefaultmidpunct}
{\mcitedefaultendpunct}{\mcitedefaultseppunct}\relax
\EndOfBibitem
\bibitem[Rom\'{a}n-P\'{e}rez and Soler(2009)Rom\'{a}n-P\'{e}rez, and
  Soler]{Roman-Perez2009}
Rom\'{a}n-P\'{e}rez,~G.; Soler,~J. {Efficient Implementation of a van der Waals
  Density Functional: Application to Double-Wall Carbon Nanotubes}.
  \emph{Physical Review Letters} \textbf{2009}, \emph{103}, 096102\relax
\mciteBstWouldAddEndPuncttrue
\mciteSetBstMidEndSepPunct{\mcitedefaultmidpunct}
{\mcitedefaultendpunct}{\mcitedefaultseppunct}\relax
\EndOfBibitem
\bibitem[Perdew et~al.(1996)Perdew, Burke, and Ernzerhof]{Perdew1996a}
Perdew,~J.~P.; Burke,~K.; Ernzerhof,~M. {Generalized Gradient Approximation
  Made Simple}. \emph{PHYSICAL REVIEW LETTERS} \textbf{1996}, \emph{77},
  3865--3868\relax
\mciteBstWouldAddEndPuncttrue
\mciteSetBstMidEndSepPunct{\mcitedefaultmidpunct}
{\mcitedefaultendpunct}{\mcitedefaultseppunct}\relax
\EndOfBibitem
\bibitem[Troullier(1991)]{Troullier1991}
Troullier,~N. {Efficient pseudopotentials for plane-wave calculations}.
  \emph{Phys. Rev. B} \textbf{1991}, \emph{43}, 1993--2006\relax
\mciteBstWouldAddEndPuncttrue
\mciteSetBstMidEndSepPunct{\mcitedefaultmidpunct}
{\mcitedefaultendpunct}{\mcitedefaultseppunct}\relax
\EndOfBibitem
\bibitem[Rocha et~al.(2006)Rocha, Garc\'{\i}a-Su\'{a}rez, Bailey, Lambert,
  Ferrer, and Sanvito]{Rocha2006}
Rocha,~a.; Garc\'{\i}a-Su\'{a}rez,~V.; Bailey,~S.; Lambert,~C.; Ferrer,~J.;
  Sanvito,~S. {Spin and molecular electronics in atomically generated orbital
  landscapes}. \emph{Physical Review B} \textbf{2006}, \emph{73}, 085414\relax
\mciteBstWouldAddEndPuncttrue
\mciteSetBstMidEndSepPunct{\mcitedefaultmidpunct}
{\mcitedefaultendpunct}{\mcitedefaultseppunct}\relax
\EndOfBibitem
\end{mcitethebibliography}

\end{document}